# Study of first-order thermal ΣΔ architecture for convective accelerometers

Olivier LEMAN    Frédérick MAILLY    Laurent LATORRE    Pascal NOUET

Laboratoire d'Informatique, de Robotique et de Microélectronique de Montpellier (LIRMM)
UMR CNRS 5506 - University Montpellier II – France
Contact: mailly@lirmm.fr

*Abstract*-This paper presents the study of an original closed-loop conditioning approach for fully-integrated convective inertial sensors. The method is applied to an accelerometer manufactured on a standard CMOS technology using an auto-aligned bulk etching step. Using the thermal behavior of the sensor as a summing function, a first order sigma-delta modulator is built. This "electro-physical" modulator realizes an analog-to-digital conversion of the signal. Besides the feedback scheme should improve the sensor performance.

## I. INTRODUCTION

Many studies report the design of open-loop convective accelerometers [1-10] but very few of them concern the conditioning electronics. Generally, a simple instrumentation amplifier is used. For long term stability and room temperature compensation, temperature control of the heater was studied by [6] and is implemented in a commercial sensor [4]. However, to our knowledge, no study deals with a *real* closed-loop architecture, where the effect of the acceleration on the detectors' temperature is compensated.

Sensor working principle is quite simple. The prototype realized at LIRMM (fig.1) using a standard CMOS technology features three polysilicon resistors embedded inside three suspended bridges. The central bridge is heated by Joule effect, and a thermal gradient is thus created in the surrounding air. The two lateral bridges are resistive temperature sensors with a Positive Temperature Coefficient (PTC). Without acceleration, both sensing bridges measure the same temperature due to symmetry reasons. Acceleration modifies the temperature profile due to free convection phenomenon along the sensitive axis which is perpendicular to the bridges. The temperature difference generated between the sensing bridges is converted into a differential output voltage by a Wheatstone bridge setup.

These sensors suffer from several weaknesses, such as:
- A quite low intrinsic sensitivity in the range of 1mV/g, which implies the design of a high-gain and low-noise amplifier.
- A non-linear acceleration-to-differential-temperature transfer function since the temperature difference between the sensors is limited by the temperature increase of the heater.
- A thermal inertia of the detectors which limits the bandwidth of the sensor to 50Hz. This is a technological limit. It was proven in previous studies that convective bandwidth higher than 100Hz could be reached through a reduction of the cavity, making thermal inertia of detectors the limiting factor [9].

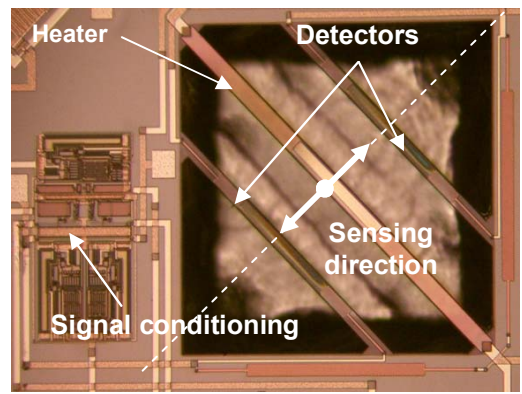

(a)

Fig. 1. Prototype micrograph

These limitations lead us to develop a closed-loop architecture in the form of an electro-physical sigma-delta modulator [11]. The sensor conditioner has a direct-digital output in the form of a bit-stream and compensates for the thermal inertia of the sensing bridges. The proposed feedback loop cancels the temperature difference between the two detectors by generating a compensation thermal power. This method uses Joule's effect, so that it can elevate the temperature of the "cold" detector, but the "hot" detector cannot be cooled this way. In order to solve this problem, the modulator slightly shifts the thermal bias point of the detectors by using a common-mode bias power which is then cancelled or increased depending on the desired action.

The first part deals with the modeling of the convective accelerometer, in order to realize behavioral simulations. In the second part we propose the electronic design of a convective accelerometer with thermal sigma-delta modulator conditioning. Finally, behavioral simulations of the proposed system are performed under the Matlab-

    

Simulink environment and the results are discussed.

As a foreword, we want to stress the point that we have experimentally validated the thermal sigma-delta modulator principle and the proposed behavioral model [10, 12 and 13], giving strength to this design study.

## II. OPEN-LOOP SENSOR MODELING

The study and modeling of the convective accelerometer sensing cell was published in previous studies [10, 13]. Using FEM simulations and characterization results, the model of fig.2 was developed.

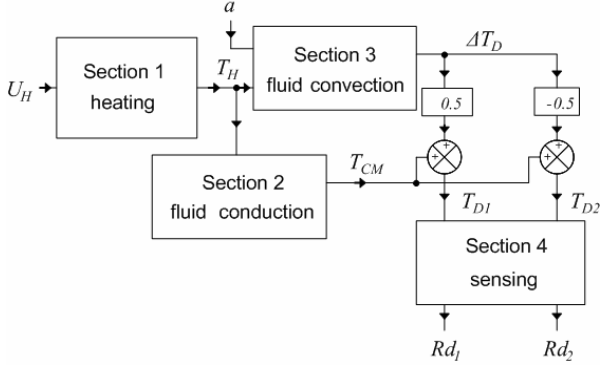

Fig. 2. Convective sensor model

Section 1 models the average temperature of the heating bridge:

$$\Delta T_H = T_H - T_A = R_{th} \times P_H \qquad (1)$$

Where $T_H$ and $T_A$ (K) are respectively the average heater temperature and ambient temperature, $R_{th}$ (K.W$^{-1}$) is the equivalent thermal resistance of the beam and $P_H$ (W) is the joule power dissipated in the heater resistance. Section 2 describes the conduction heat transfer, which is responsible for the common-mode temperature of the detectors. This model is based on a cylindrical geometry approximation and the common-mode temperature at a distance $r$ from the heater is ruled by expression 2:

$$T + \gamma \frac{T^2}{2} = \frac{\left[(T_H - T_A) + \frac{\gamma}{2}(T_H^2 - T_A^2)\right]}{\ln\left(\frac{r_2}{r_1}\right)} \cdot \ln\left(\frac{r}{r_1}\right) + T_H + \gamma \frac{T_H^2}{2} \qquad (2)$$

Section 3 expresses convective transfer from which depends the output signal. The temperature profile established in section 2 deforms, producing a temperature difference between the detectors $\Delta T_D$. This temperature difference has been found to be proportional to the Grashof number [1] for $\Delta T_D \ll (T_H - T_A)$:

$$\Delta T_D = \frac{S.Gr}{1+\tau p} = S \cdot \frac{a\beta\rho^2(T_H - T_A)l^3}{\mu^2} \cdot \frac{1}{1+\tau p} \qquad (3)$$

with $S$ (in K) a fitting coefficient (extracted from FEM analysis and experimental results) which represents the system sensitivity, a (m.s$^{-2}$) the acceleration, $\rho$ (kg.m$^{-3}$) the gas density, $\beta$ (K$^{-1}$) the gas coefficient of expansion, $\mu$ (kg.m$^{-1}$.s$^{-1}$) the gas viscosity, $l$ (m) a linear dimension related to the cavity volume and geometry and $\tau$ (s) the time constant of the fluid ($\tau$ = 0.5 … 12ms, depending on the cavity geometry and dimensions [9]).

Finally, section 4 implements detectors transfer function taking into account the temperature sensitivity of polysilicon ($TCR = 9.10^{-4}$/K) and the thermal time constant of the detecting bridges ($\tau_D = 3.3$ms):

$$Rd_i = Rd_{i,nom}\left(1 + TCR\left(T_{CM} \pm \frac{1}{2}\Delta T_D\right)\frac{1}{1+\tau_D p}\right) \qquad (4)$$

The resistance variations of the PTC sensors are converted into a differential output voltage by means of a Wheatstone bridge with a sensitivity $S_{wheat}$ (V/K):

$$S_{Wheat} = \frac{Vdd}{4}\frac{TCR}{1+TCR.T_{MC}} \qquad (5)$$

The Johnson noise of the detectors is finally added to the model since it has been identified as the limiting factor for the sensor resolution. In our prototype, the heating power is 42mW, heater temperature is 720K and detectors common-mode temperature is 423K. The convection transfer function can thus be linearized as $\Delta T_D = a.K_{MEMS}$ with sensitivity $K_{MEMS}$ = 1.53K/g. $K_{MEMS}$ is an empirical, composite parameter including the parameters of expression 3. Wheatstone bridge's Johnson noise ($4k_bT_{CM}R$) generated by the 50kΩ resistors is 34nV/√Hz over a 6MHz bandwidth limited by the 1pF comparator input capacitance. Accelerometer sensitivity is 1.55mV/g, and its intrinsic resolution is 98μgrms over the 1-20Hz bandwidth.

In order to study the conditioning electronics, a small signal model of the sensor was derived (fig. 3). Such a simplified model is suitable for fast behavioral simulations under the Matlab-Simulink framework.

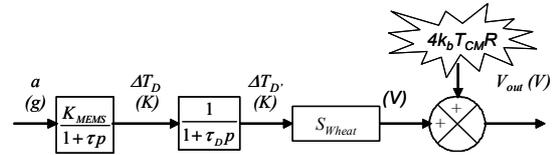

Fig. 3. Small-signal model of the sensor

This model was used in a previous study of open-loop amplifier architectures to show that a Wheatstone bridge bias modulation together with a synchronous detection scheme would allow to cancel the low-frequency noise of a standard CMOS amplifier and to reach resolutions close to the intrinsic performance of the convective accelerometer cell [12].

## III. CLOSED-LOOP SENSOR ARCHITECTURE

The convective sensing element studied here is an evolution of the previous version presented in part 2. Two temperature detectors are added in the suspended bridges in order to create a Wheatstone bridge with 4 sensitive resistors (fig. 4-a). The resulting sensor features twice the original $S_{Wheat}$ sensitivity and then twice the previous intrinsic signal-to-noise ratio. Considering the operating conditions stated in part 2, the sensitivity reaches now 3.1mV/g; the intrinsic noise of 11µg/√Hz leads to a 49µgrms resolution over the 1-20Hz bandwidth.

In order to allow a thermal feedback of the suspended bridges, self-heating of the temperature detectors due to the biasing of the Wheatstone bridge is now taken into account in the model, resulting in a slight shift of the common mode temperature of the two suspended bridges. Moreover, we noticed that this self-heating interestingly falls in the same range as the power necessary to compensate for the convection phenomenon. The thermal feedback is then simply achieved by shorting the sensitive resistors using MOS switches (fig. 4-a). Table 1 gives the power dissipation in the two suspended bridges as a function of the MOS switches command. When both feedback commands are low, the Wheatstone bridge is regularly biased, and the power dissipation is $P_0 = V^2/(2R(T_{MC})) = 225µW$ in each suspended bridge. When feedback_1 is high, resistors in bridge 2 are shorted and resistors in bridge 1 are biased at $V_{dd}$. Power dissipation is then null in bridge 2 and $P_{max} = 2V^2/(R(T_{MC})) = 900µW$ in bridge 1. Feedback_2 activation generates a complementary power dissipation effect. These two feedback paths are used to compensate the temperature difference generated by convection (*i.e.* by the acceleration).

TABLE 1
Control signals and corresponding power dissipation in sensing bridges

| command | | dissipation | |
|---|---|---|---|
| feedback_1 | feedback_2 | bridge 1 | bridge 2 |
| 0 | 0 | $P_0$ | $P_0$ |
| 1 | 0 | $P_{max}$ | 0 |
| 0 | 1 | 0 | $P_{max}$ |

This principle requires a 2-phase time scheduling and fig. 4.b. presents the corresponding chronograms:
- phase Φ1, the sensor is in reading mode. Both feedback paths are disabled, and the Wheatstone bridge under regular bias delivers the error signal $V_{out}$. If $V_{out}$ is negative, it means that the two detectors embedded on bridge 2 are the warmer and then the comparator output is low. If it is positive, the two detectors in bridge 1 are the warmer and the comparator output is high.
- phase Φ2, the sensor is in the feedback mode, thus one of the two feedback paths is enabled according to the comparator output. For example, if the comparator output is low, feedback_1 is enabled: power dissipation is then null in bridge 2 and equal to $P_{max}$ in bridge 1. During this phase, the Wheatstone bridge output is an invalid +/- $V_{dd}$.

The longer $T_{comp}$ is, the higher compensation power $P_{comp}$ is generated, according to the $P_{comp} = α P_{max}$ relationship. Sensor full-scale is adjusted through the duty-cycle $α$ of feedback pulses. A null acceleration entails that each bridge is heated by an average $P_{comp}/2$ power which slightly shifts the thermal common-mode of the detecting bridges.

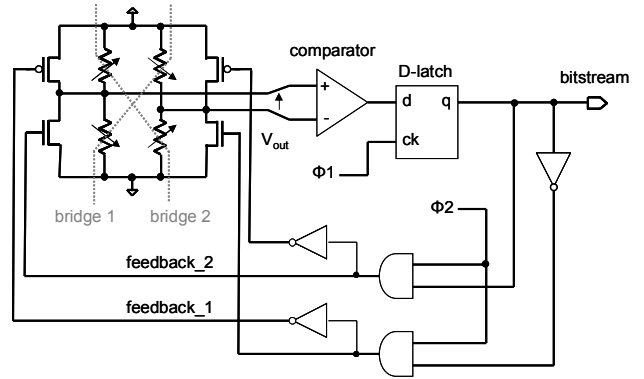

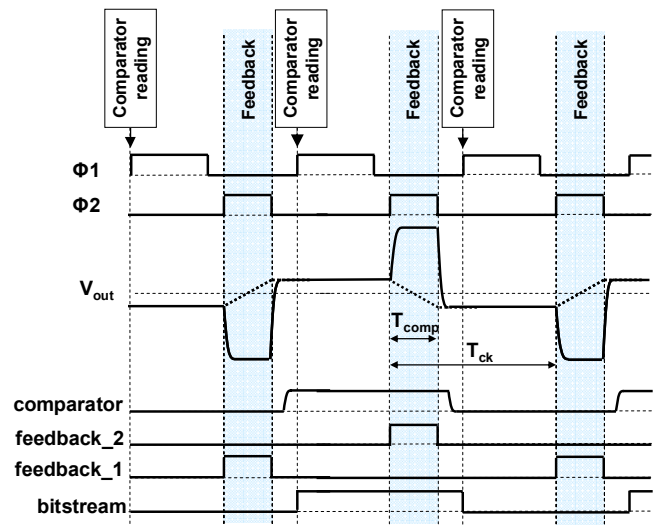

(b)

Fig. 4. Modified sensor architecture (a) and chronograms (b)

IV. CLOSED-LOOP SENSOR SIMULATION

First the small signal model of the sensor (fig.3) is modified to include the power-feedback scheme of the sigma-delta modulator (fig.5). This is achieved by converting the temperature variations of the bridges into thermal power using the characterized thermal resistance of the detecting bridges, $R_{th} = 10^4$K/W. Consequently, system input becomes a thermal power $ΔP_D$ resulting from acceleration:

$$ΔP_D = \frac{K_{MEMS}}{R_{th}} \frac{a}{1+τp} \qquad (6)$$

In this first order thermal sigma-delta modulator, the electro-thermal response of the detecting bridges replaces the integrator. This thermal filter is a first-order low-pass with a cut-off frequency of 48Hz, so that it has a DC gain of $10^4$K/W in the DC-48Hz part of the spectrum, thus acting as a leaking integrator.

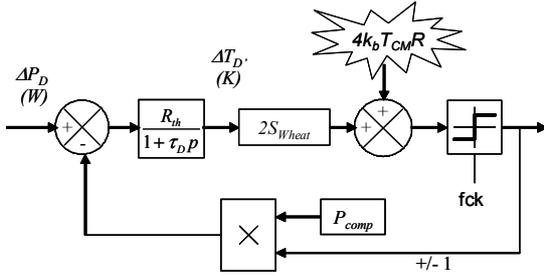

Fig.5. Sigma-delta modulator simulation model

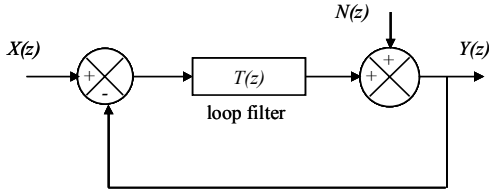

Fig. 6. Linearized model of the sigma-delta modulator

In order to study the impact of this non-ideal integrator, we use a linearized model of the modulator (fig. 6) where T(z) is the z-transform of the loop filter transfer function and N(z) is the quantization noise, under the white-noise assumption. The Johnson noise is not considered in this model.

Quantization noise transfer function Y(z)/N(z) is :

$$NTF(z) = \frac{1}{1+T(z)} \quad (7)$$

Considering an ideal integrator as the loop filter, the NTF equals $1-z^{-1}$. Hence, quantization noise is shaped over the full spectrum. Now if we consider the thermal filter which transfer function in the z-domain writes:

$$T(z) = \frac{z^{-1}}{1 - \alpha \cdot z^{-1}} \quad (8)$$

with $\alpha = e^{-\frac{Te}{\tau_D}}$ close to 1, the NTF becomes $1 - \alpha \cdot z^{-1}$ [14]. This NTF becomes flat in the DC-48Hz part of the spectrum, meaning that the quantization noise is unshaped in this part of the spectrum (fig. 7-b).

Transient behavioral simulations of the system were carried-out in Matlab-Simulink. We studied the performance of the architecture using a spectral analysis of the output bit-stream (fig. 8).

If we compare results of figures 7 and 8, we notice that the behavior of the simulation model is close to the prediction of the analytical model in the low-frequency spectrum, the low-pass thermal loop filter leads to a flat quantization noise from 1 to 48Hz (fig.8-a). Resolution is 1.48mg over the 1-20Hz bandwidth, whereas the same modulator with an ideal integrator (fig.8-b) would have a 0.26mg resolution. Moreover, a slight distortion appears. This phenomenon is common for first-order modulators, it corresponds to the third harmonic and its amplitude is below 0.1% of the main peak.

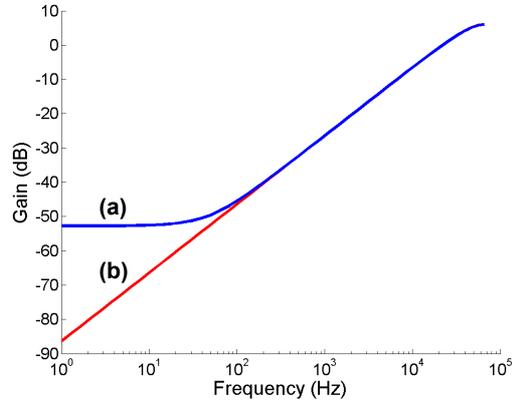

Fig. 7. Noise transfer functions for an ideal integrator (b) and for the thermal integrator (a) @ $f_{ck}$ = 131kHz.

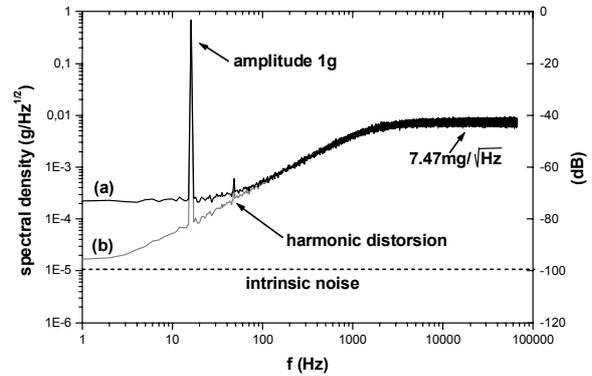

Fig. 8. FFT of the bit-stream for the sigma-delta modulator model (a), same modulator with an ideal integrator (b), $f_{ck}$ = 131kHz, ±2g full-scale, input : 16Hz sinusoidal acceleration, amplitude 1g

However, figure 8 shows that quantization noise is unshaped in the upper part of the spectrum. Above 1.6kHz, the power of the error signal falls below the Johnson noise floor of the Wheatstone bridge. This results in a random output of the 1-bit quantizer. In the end, oversampling simply leads to a quantization noise spreading. In absence of shaping from 1.6kHz to $f_{ck}/2$ (e.g. about 98% of the spectrum), more than 99% of the quantization noise power is unshaped. We can have a good estimation of the quantization noise spectral density in the 1.6kHz-($f_{ck}/2$) spectrum, neglecting the 48Hz-1.6kHz first order noise shaping. Assuming that the total bit-stream power is constant, the quantization noise should be given by:

$$\frac{\sqrt{S_b^2 - S_a^2}}{\sqrt{f_{ck}/2}} = 7.31 mg/\sqrt{Hz} \quad (9)$$

where $S_b$ = 2$g_{RMS}$ is the bit-stream signal power, and $S_a$ =

$1/\sqrt{2}g_{RMS}$ is the acceleration signal.

Simulations of the figure 8 give 7.47mg/√Hz. This proves that the 48Hz-1.6kHz 1st order shaping has little effect on the overall quantization noise shaping.

Figure 9 presents the signal transfer functions of the two modulators. As a consequence of the finite DC gain of the thermal, the DC gain of the electro-thermal modulator is 0.3dB lower than the theoretical gain. Their -3dB cut-off frequencies are both about 1.6kHz. Above this frequency it was shown that the Johnson noise of the Wheatstone bridge disturbs the modulator operating. It's likely that this phenomenon limits also the modulator bandwidth.

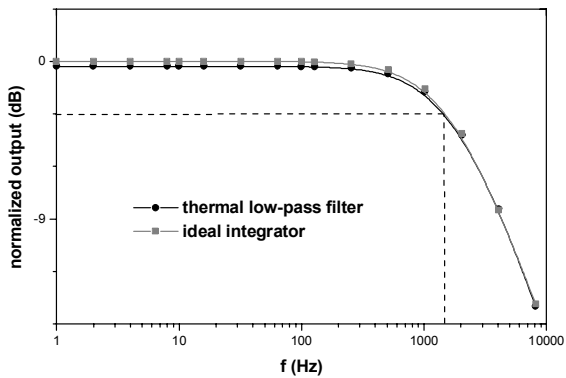

Fig. 9. Signal transfer functions of the modulators, $f_{ck}$ = 131kHz, ±2g full-scale

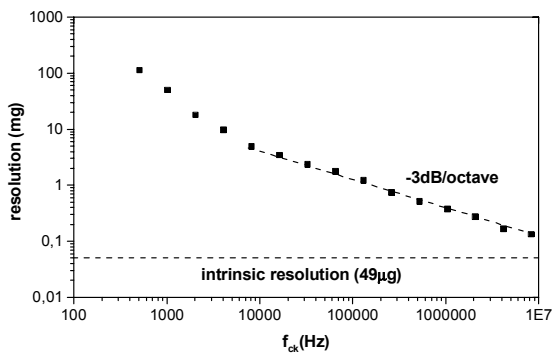

Fig. 10. Resolution (BW=1-20Hz) versus modulator clock frequency

Finally, the same system was simulated for various clock frequencies $f_{ck}$ and the corresponding resolutions were computed over the bandwidth 1-20Hz. The results, plotted on figure 10, show that for clock frequencies higher than 10 kHz, the signal-to-noise ratio increases at a 3dB/octave rate. This poor improvement comes from the quantization noise spreading due to the absence of quantization noise shaping above 1.6kHz.

A 132μg resolution was simulated with a 8.4MHz clock frequency. Higher clock frequencies resulted in bitstreams too large to be analyzed with Matlab. From the -3dB/octave SNR increase, a modulator with a clock frequency of about 100MHz would reach the intrinsic resolution of the convective sensor; say 49μg over the 1-20Hz bandwidth.

This result suggests the design of a second order sigma-delta modulator architecture, expecting a lower distortion and lower frequency clocks through a better noise shaping.

## CONCLUSION

This study reports an original smart-sensor architecture for thermal convective accelerometers based on the thermal sigma-delta modulator principle. We consider a first order sigma-delta architecture. Using the side effects of self heating and thermal inertia of detectors, most of the analog signal processing is performed in the thermal domain, drastically reducing the amount of analog electronics. Simulation results show that a high modulator clock frequency is required to reach the intrinsic noise of the sensor, motivating the design of a higher order modulator.